\documentclass[%
preprint,
 amsmath,amssymb,
 aps,
 pre,
]{revtex4-2}

\usepackage{graphicx}
\usepackage{dcolumn}
\usepackage{bm}%
\usepackage{color}
\usepackage{braket}
\begin{document}

\preprint{Boston Studies in the Philosophy and History of Science, Manuscript}

\title{~\\ Hydrodynamically Inspired Pilot-Wave Theory:\\ An Ensemble Interpretation}

\author{Yuval Dagan}%
 \email{yuvalda@technion.ac.il}
\affiliation{%
 Faculty of Aerospace Engineering, Technion - Israel Institute of Technology, Haifa, 320003, Israel.\\ \\
}%


\begin{abstract}
This chapter explores a deterministic hydrodynamically-inspired ensemble interpretation for free relativistic particles, following the original pilot wave theory conceptualized by de Broglie in 1924 and recent advances in hydrodynamic quantum analogs. 
We couple a one-dimensional periodically forced Klein-Gordon wave equation and a relativistic particle equation of motion, and simulate an ensemble of multiple uncorrelated particle trajectories.
The simulations reveal a chaotic particle dynamic behavior, highly sensitive to the initial random condition.
Although particles in the simulated ensemble seem to fill out the entire spatiotemporal domain, we find coherent spatiotemporal structures in which particles are less likely to cross. These structures are characterized by de Broglie's wavelength and the relativistic modulation frequency $k c$.
Markedly, the probability density function of the particle ensemble correlates to the square of the absolute wave field, solved here analytically, suggesting a classical deterministic interpretation of de Broglie's matter waves and Born's rule.
\end{abstract}

\maketitle

\newpage

\section{Introduction}
In 1924, de Broglie proposed that particles may be associated with an intrinsic clock oscillating at the Compton frequency. 
De Broglie envisaged a particle as a localized wavefield guided by a pilot wave, exchanging rest-mass energy with field energy~\cite{de1924recherches}.
Interactions between the particle generating the wave field and, in turn, the wavefield guiding the particle constituted de Broglie's realistic picture of matter~\cite{de2012heisenberg}.

The particles in de Broglie's theory were conceptualized as a localized yet infinite, spatially decaying field guided by relativistic phase waves. 
Relativistic considerations, and in particular the notion of \textit{Harmony of Phases} associating the particle to its guiding wave, were imperative in de Broglie's theory: \\

\textit{``I think the theory of relativity plays a major role much greater than most people usually think in the basic ideas of wave mechanics and that if one wants to really understand its origins, one has to come back to relativistic considerations..."}, Louis de Broglie, 1967.\\

This note from an interview (translated here from French) reveals de Broglie's view on the fundamental importance of relativistic motion in the dynamics of particles. Thus, relativistic considerations could  also play an important role in the hydrodynamic pilot-wave analogy~\cite{dagan2023relativistic}. 
In de Broglie's studies, matter waves were generally realized as monochromatic plane waves or at least quasi-monochromatic, as appeared in his later notes~\cite{de1970reinterpretation}. 
However, de Broglie neither defined a specific guiding wavefield in his theory nor suggested a mechanism for the generation of such waves (although he did suggest several candidates for such waves, including the Klein-Gordon equation and the Dirac equation~\cite{de1970reinterpretation}).

The astounding success of Born and Heisenberg in mathematically describing the statistical nature of quantum mechanics, and perhaps de Broglie's inability to finish his incomplete program, eventually led to generally discarding the realistic pilot-wave view of quantum particle dynamics.  
In retrospect, in an attempt to find a realistic picture of matter, de Broglie's assumptions may have seemed too simplified and thus incomplete.
Whether the pilot wave theory conceptualized by de Broglie describes a realistic picture of matter or not, there is a reason to believe that it should constitute more complex two-way coupled particle-wave interactions that were not realized in de Broglie's program. 

Nonlinear particle-wave interactions, similar to ideas conceptualized by de Broglie's later publications, are frequently encountered in fluid dynamics. Since the introduction of the Madelung transformation~\cite{Madelung1926} of the Schr\"odinger equation to a fluid mechanical system of equations, multifold attempts to realize quantum particle dynamics relying on fluid mechanics principles have been proposed. De Broglie-Bohm theory~\cite{Bohm1952}, Nelson's theory~\cite{Nelson1966}, and the Stochastic Electrodynamics (SED)~\cite{boyer1975random, de2015emerging} have raised the possibility that underlying unknown statistical mechanisms - the so-called `hidden variables' - govern the dynamics of quantum particles giving rise to the expected quantum statistical signature. 

However, de Broglie's original picture of matter was neither a hidden variable approach nor statistical but rather a classical deterministic one.
It is, therefore, instructive to follow de Broglie's deterministic views and review some of the recent advances in deterministic hydrodynamic quantum mechanical analogs. 
One of the most successful analogies to quantum mechanics was found by Couder and Fort, who experimentally observed millimetric oil droplets bouncing over a vibrating bath that remarkably feature the statistical behavior of many quantum mechanical systems~\cite{Couder2005, Couder2006, BushOza2020}.
In this hydrodynamic quantum analogy (HQA), droplets interact in resonance with a quasi-monochromatic wavefield they generate and exhibit a self-propelling mechanism. This analog has extended the range of classical physics to include many features previously thought to be exclusively quantum, including tunneling \cite{eddi2009unpredictable,hubert2017self,nachbin2017tunneling,tadrist2020predictability}, Landau levels~\cite{fort2010path,harris2014droplets,oza2014pilot}, quantum harmonic oscillator~\cite{perrard2014self,durey2020hydrodynamic}, the quantum corral~\cite{harris2013wavelike,gilet2014dynamics,gilet2016quantumlike,saenz2018statistical,cristea2018walking}, the quantum mirage~\cite{saenz2018statistical}, and Friedel oscillations~\cite{saenz2020hydrodynamic}.
Remarkably, Couder was able to demonstrate in his hydrodynamic analogy a mechanism for single-particle diffraction \cite{couder2006single}.

The ability of a classical dynamic system to resolve quantum mechanical problems raises the question: Are we able to conceptualize a similar deterministic mechanism on the microscopic (quantum mechanical) scale? 
In this chapter, we shall address this question theoretically using hydrodynamic considerations in an attempt to realize deterministic particle dynamics on the quantum scale. 

Several authors have attempted to formulate a form of quantum dynamics based on insights from HQA~\cite{Andersen2015, Borghesi2017, Drezet2020}.
Andersen et al.~\cite{Andersen2015} explored the behavior of a dynamical system in which a particle locally excites a waveform satisfying Schr\"odinger's equation, which then moves in response to gradients of the phase of that field. Orbital quantization was shown to arise along with an analog to the Bohr-Sommerfeld quantization rule; Borghesi~\cite{Borghesi2017} proposed an elastic pilot-wave model wherein a point particle moves within a non-dissipative elastic substrate. The coupled dynamics of the particle and elastic medium in Borghesi's study are governed by a modified Klein-Gordon equation. Shinbrot~\cite{Shinbrot2019} examined the Klein-Gordon equation with an oscillatory potential as a model for a quantum particle emitting and absorbing pilot waves. He concluded that bound state solutions with half-integer spin exist provided the particle rest mass oscillates in time, which aligns with de Broglie's physical picture~\cite{deBroglie1970}. 

Recently, we developed a hydrodynamically-inspired quantum theory~\cite{dagan2020hydrodynamic}, a theoretical model of relativistic quantum dynamics inspired by de Broglie's pilot wave theory. 
In this framework, the particle is assumed as a localized - yet infinite - oscillating disturbance, externally forcing a Klein-Gordon wave equation. 
A relativistic dynamic equation couples the motion of the localized particle to the wave.
Using this deterministic framework, several features of quantum mechanics are revealed, associated with inline oscillations that correspond to the relation $p=\hbar k$, realized through interactions with the wave field. Notably, particle speed modulations are averaged at the de Broglie wavelength and modulated by the relativistic frequency $k c$.
Although the nonlinear system of free particles in this framework is chaotic in nature, and inline oscillations may be characterized by multiple oscillation modes, de Broglie wavelength is the most pronounced and may be realized as \textit{quasi-monochromatic} modulation of the particle motion, as also suggested by de Broglie. 

Excitation of motion and the waveform of the same framework at non-relativistic speeds were examined by Durey and Bush~\cite{Durey2020b}, who revealed the wave generation and self-propelling mechanism for the coupled wave-particle system, and provided analytical validity to our subsequent work on the hydrodynamic pilot-wave theory. 

To further explore the extent to which a classical hydrodynamic theory may be realized as a viable interpretation of de Broglie's theory and relativistic quantum dynamics, and in particular, the possibility of such a theory to account for de Broglie wavelength, a fully classical non-relativistic dynamic system was also considered~\cite{dagan2023relativistic}.
This is in contrast to our previous study~\cite{dagan2020hydrodynamic}, where a relativistic dynamic equation is introduced to properly satisfy a Lorentz covariant formulation. 
However, this formulation closely correlates to the hydrodynamic analog and allows the isolation of the role of classical wave mechanics in producing relativistic quantum signatures. 

In the present study, we extend the discussion on the relativistic pilot wave framework developed by Dagan and Bush~\cite{dagan2020hydrodynamic} by simulating multiple uncorrelated particle simulations and analyzing the statistics emerging from the ensemble of particle trajectories. 

\section{Hydrodynamically-inspired quantum theory}
Several options for realizing particle-wave interactions may be postulated. 
We follow our previous study~\cite{dagan2020hydrodynamic},
where the pilot waves are generated by the oscillating particle and evolve according to a one-dimensional KG equation:
\begin{equation}\label{eq:KG}
	\frac{\partial^2 \phi}{\partial t^2} - c^2 \frac{\partial^2 \phi}{\partial x^2} + \omega_c^2 \phi = 
	\epsilon_p f(t) \delta_a(x-x_p)~, 
\end{equation}
Here, $\varphi$ is a real wave, $\omega_c$ is the Compton frequency, and $c$ is the speed of light.
The term on the RHS of equation (\ref{eq:KG}) represents an external forcing of the wave-field localized about the particle location 
$x_p$, where $f(t)=\sin(2\omega_c t)$, $\epsilon_p$ is a constant and
$\delta_a = \frac{1}{|a|\sqrt{\pi}}e^{-(x/a)^2}$  is a modified delta function that serves to localize the driving oscillation to 
the vicinity of the particle location. The parameter $a$ defines the width of the modified delta function and the extent of the particle's influence on the wave field. 
Here, as in~\cite{dagan2020hydrodynamic}, $a=\lambda_c/2$, where $\lambda_c = h/(mc)$ is the Compton wavelength. 
Hence, the Gaussian perturbation has a characteristic width of $\lambda_c$. $\epsilon_p$ is the forcing amplitude, and $f(t)$ is chosen as a harmonic function with twice the Compton frequency, $f(t) = sin(2\omega_c t)$.
Inspired by HQA and de Broglie's guidance equation, a trajectory equation can be written in which the momentum is determined by the gradients of the waves such that 
\begin{equation}\label{eq:traj_HQFT1}
\gamma \dot{x}_p = -\alpha \frac{\partial \phi}{\partial x'}_{x=x_p}~,
\end{equation}
where $\dot{x}_p$ is the particle velocity, and $x'$ denotes the particle location after proper Lorentz translation from the particle frame of reference to the stationary one. Note that this translation ensures the equation of motion is relativistically covariant yet not invariant as the Klein-Gordon equation. 
$\alpha$ is a free coupling parameter between the particle and its waves, and $\gamma$ is the Lorentz boost factor.
The LHS represents the relativistic momentum, divided by the rest mass $m_0$, which is absorbed into the coupling parameter, $\alpha$.  
Notably, when a classical particle is coupled to periodic and oscillating flows, clustering of particles may occur, which may be viewed as quantization of their probability distribution~\cite{dagan2017similarity, dagan2017particle}.     

We shall now solve the coupled equations numerically for a single free particle.
The coupled Klein–Gordon equation and the equation of motion are discretized using finite differences. An explicit finite-difference method was derived to solve the Klein–Gordon wave equation, and a Runge–Kutta scheme was employed to advance in time the nonlinear guidance equation.
More details on the numerical method and numerical parameters may be found in~\cite{dagan2020hydrodynamic}.

\begin{figure}
    \begin{center}        
	\includegraphics[width=1\textwidth]{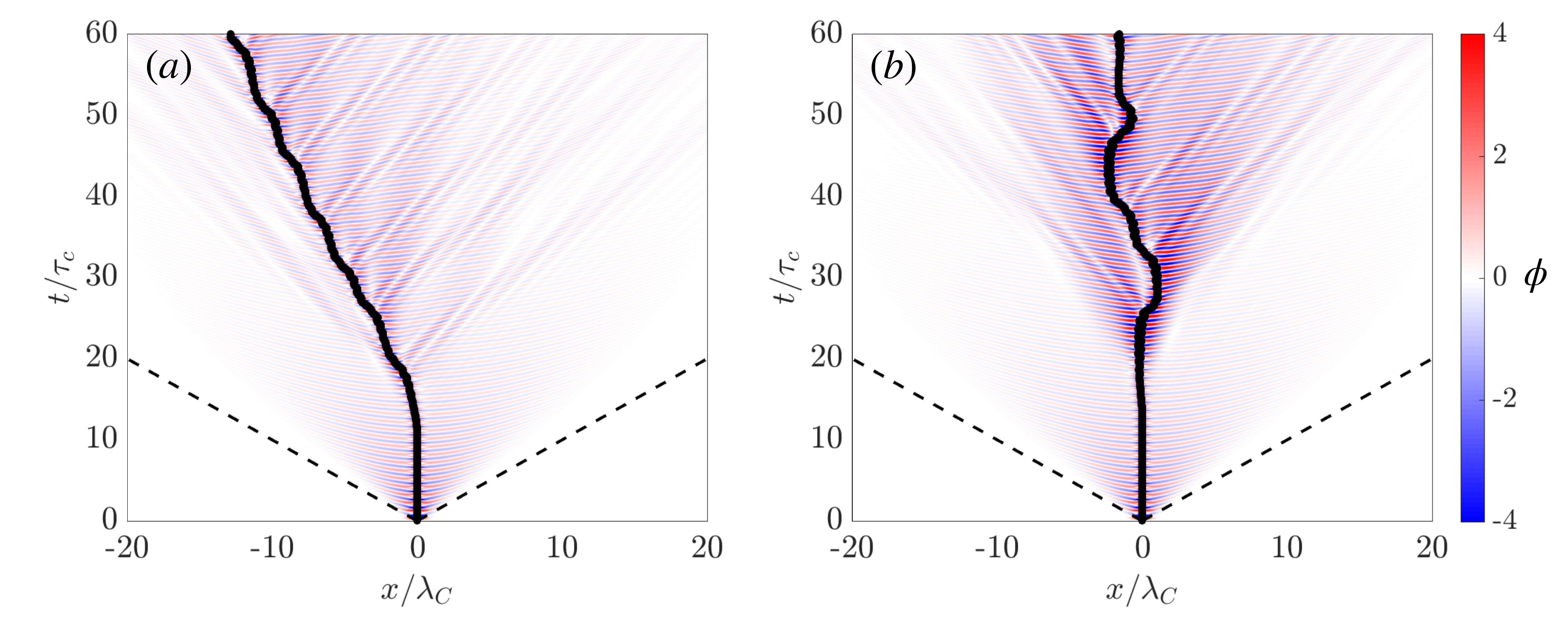}
    \caption{Pilot-wave dynamics of two free particles demonstrated in (a) and         (b) for two different initial conditions. The spatiotemporal map illustrates the evolution of the normalized pilot-wave field, $\phi$ (color map), generated by the trajectory of a free particle (indicated in black). The coupling constant is set to $\alpha = 0.045$, corresponding to a quasi-steady speed of approximately $\beta = 0.25$, as calculated from the wavelength in the particle vicinity. The particle's light cone is indicated by the dashed black line with slope $\beta=1$.}
    \label{fig:hqft}
    \end{center}
\end{figure}

Figure~\ref{fig:hqft} demonstrates the typical spatiotemporal evolution of the wavefield and particle dynamics for two different cases, which differ only by the initial condition imposed on the particle; in both cases, an initial stage of motion may be observed, at which the particle seems stagnant during approximately the first ten Compton periods.
The particle oscillates at twice the Compton frequency, exciting localized standing waves. To break the symmetry and initiate motion, an initial random wave perturbation is applied, approximately four orders of magnitude smaller than the characteristic wave amplitude generated by the stagnant particle. This perturbation initially causes small oscillations about its initial position. 

In fact, since the particle is initially deflected randomly, inline oscillations about $x/\lambda_C=0$ occur but are too small to be observed in the scale of the figures. However, symmetry breaking in both the wavefield and particle motion is apparent between $t/\tau_c=10$ and $t/\tau_c=20$. 
In figure~\ref{fig:hqft}a, the particle locks into a quasi-steady motion, as also reported in our previous study~\cite{dagan2020hydrodynamic}. In figure~\ref{fig:hqft}b, however, a more chaotic motion is apparent, for which we cannot determine a characteristic quasi-steady speed, although the same slope, i.e., speed of $\beta = 0.25$, may be observed locally. 
In this case, as in any other simulated trajectory, the particle will eventually lock to the quasi-steady speed if the simulation is integrated for a long enough time. 

This peculiar behavior reveals the complexities that may emerge from the highly nonlinear coupled system comprising our fully deterministic model. 
On the one hand, the evolution of the wavefield and, in turn, the particle motion is highly sensitive to initial conditions and reflects chaotic dynamics.  
On the other hand, the particle locks into inline oscillations, where coherent waveform may be observed. 

Notably, as previously reported~\cite{dagan2020hydrodynamic}, the instantaneous momentum of the particle at any time corresponds to $p=\hbar k$, where $k$ is measured locally in the vicinity of the particle. Farther away from the particle location, the dispersion of waves and superposition with waves reflected from previous changes of motion affects the wavelength, and the correlation to the particle momentum is less significant.
This was confirmed for several low and high speeds along the trajectory shown in figure~\ref{fig:hqft}b.

The complex dynamics show promising features for a deterministic hydrodynamic interpretation of quantum mechanics, including a mechanism for excitation of motion, rendering the system unstable, and quantized momentum at the de Broglie wavelength.
To further investigate the ability of this model to account for quantum statistics, multiple trajectories are simulated and analyzed in the following section.

\section{Ensemble of particle trajectories} 
\begin{figure}
    \begin{center}        
	\includegraphics[width=1\textwidth]{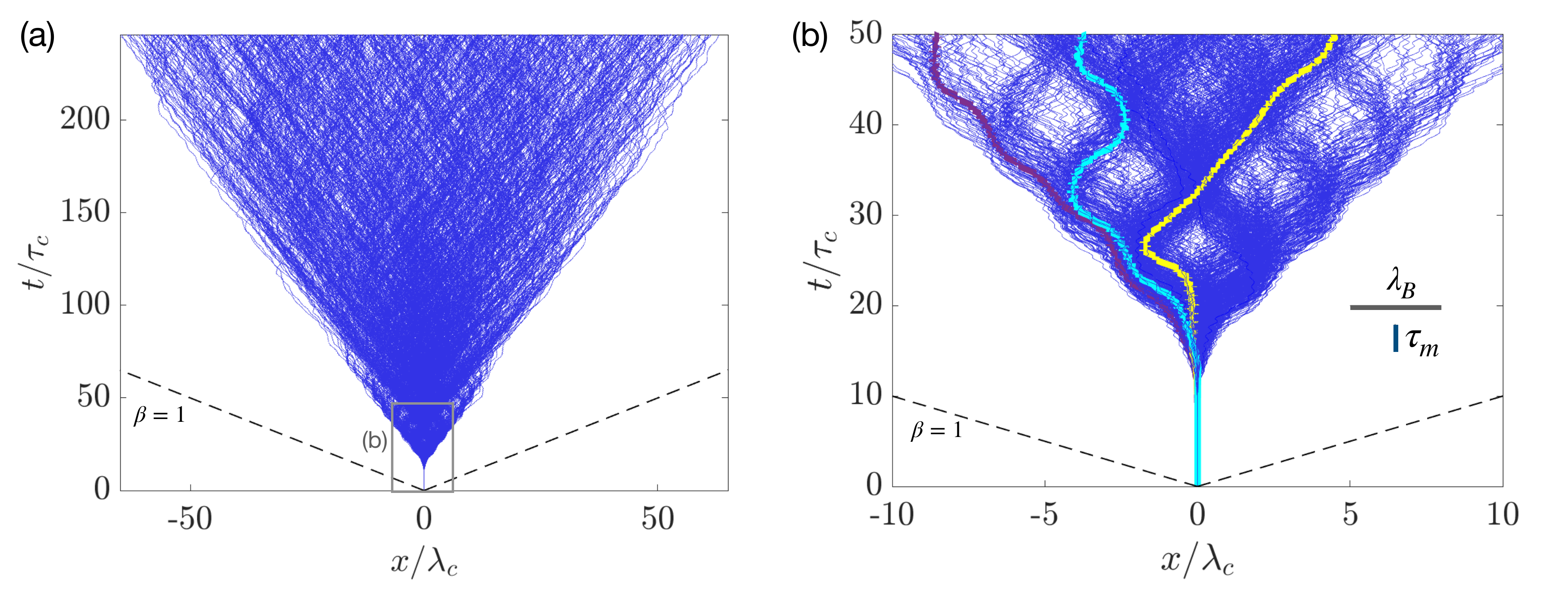}
	\caption{(a) Ensemble of 1000 particle trajectories simulated at the same settings as in figure~\ref{fig:hqft}. (b) A close-up view of the initial stages of evolution. Blue lines show all trajectories; Three particular trajectories highlighted in colors illustrate different scenarios for the particle spatiotemporal path. Dashed lines mark the speed of light cone of $\beta=1$. Reference scales for de Broglie wavelength $\lambda_B$ and relativistic modulation $\tau_m$ are shown here in horizontal grey and vertical blue lines, respectively.}
	\label{fig:ensemble}
    \end{center}
\end{figure}

We proceed by computing multiple particle trajectories. An ensemble of $1000$ overlapping trajectories is presented in figure~\ref{fig:ensemble}a, and a close-up view in figure~\ref{fig:ensemble}b under the same conditions presented in figure~\ref{fig:hqft}. 
Here, each line represents a distinct simulation, which differs from all others only by applying an initial condition of random perturbation. 

The three highlighted trajectories of figure~\ref{fig:ensemble}b illustrate different scenarios for the spatiotemporal particle evolution. Dashed lines mark the light cone with respect to the initial position of all simulations at $x/\lambda_c=0$.

In the first stages of evolution, all trajectories start at the exact spatial location at $x / \lambda_c=0$. This location is sustained for some time until symmetry braking is apparent. 
After about ten Compton periods, trajectories seem to diverge, and particles fill the space. 

As in figure~\ref{fig:hqft}, each particle is characterized by inline oscillations corresponding to the local wave field and de Broglie wavelength.
When viewed at a length scale around $\lambda_B/\lambda_c\approx 100$, particle trajectories seem randomly scattered, and as time proceeds, we observe a more spatially uniform distribution of particles.
However, a closer look at the first stages of the ensemble evolution (figure~\ref{fig:ensemble}b) reveals a clear spatial preference for the spatiotemporal path of the particle.
Coherent structures of the ensemble appear through frequent changes in the mean particle direction, resulting in spatiotemporal voids in which particles are less likely to cross.
Notably, these voids have a typical length scale comparable to de Broglie wavelength $\lambda_B$, and time scale $\tau_m$ (scales shown for reference in figure ~\ref{fig:ensemble}b).


It is interesting to note three distinct trajectories (highlighted):
The purple trajectory attains the quasi-steady speed of $\beta = 0.25$ and follows a path similar to the envelope of the ensemble;
the trajectory in cyan appears to change directions after approximately $30$, and then $40$ Compton periods, while the yellow trajectory appears to lock into a quasi-steady mean speed after abruptly changing directions around $25$ Compton periods.
Note that the purple and cyan trajectories start with roughly the same path until diverting into two distinct routes.

And so, it seems that particle-wave interactions create spatiotemporal unstable nodes in which a bifurcation appears only when observed as an ensemble.
In fact, all trajectories are characterized by the same quasi-steady speed observed in figure~\ref{fig:hqft}, and we may also assume they all interact with the local wave field and follow de Broglie's relation $p=\hbar k$ (this was confirmed for multiple points in the Spatio-temporal map for different local velocity changes - see also~\cite{dagan2020hydrodynamic}).

The emergence of spatiotemporal quantization of the ensemble reveals the unique particle-wave duality in this classical analog. 
While the dynamics of a free particle are governed by inline oscillations characterized by de Broglie wavelength, their interactions with the wavefield include both chaotic and coherent structures. 
Although single particle trajectories are perfectly deterministic, multiple successive particle trajectories form a clear temporally evolving diffraction pattern.

One can now wonder how the statistics of this ensemble correspond to Born's rule of the standard Copenhagen interpretation, in which the absolute real-valued wave function represents the probability of determining particle position.  
In the following sections, we shall show in a more quantitative manner that our deterministic framework indeed captures, in general, the correct statistical form of the standard quantum mechanical representation.  

\section{Quantum statistics of a classical ensemble} 
We now turn to look at the statics of the ensemble to better understand its evolution and viability in reproducing quantum statistics. 

Due to a relatively large number of simulated trajectories, we may extract the probability density function (PDF) of the particle location at each time step.  
The time evolving PDF is demonstrated in figure~\ref{fig:KG_an_sim}a. 
Initially, the PDF takes the form of a Gaussian (due to its normal distribution kernel), representing all particles localized at $x/\lambda_c=0$. One may notice that it then evolves into a spatial structure similar to that of a wave field responding to an initial perturbation. 
To test this hypothesis, we explore the response of the Klein-Gordon wave field to an initial Gaussian perturbation of the same characteristics as determined from the simulated particle PDF. 

\subsection{Wave response to a Gaussian disturbance}

In order to analytically resolve the underlying wave field, we consider the response of the unsteady Klein-Gordon equation in one dimension to an initial spatial disturbance $\psi_0(x)$
\begin{align}\label{eq:kg}
	\frac{\partial^2 \psi}{\partial t^2} - c^2\frac{\partial^2 \psi}{\partial x^2} + \omega_c^2\psi = 0 ~, \\ \nonumber
	\psi(x,0) = \psi_0(x)
\end{align}

We shall now define separation parameters and assume
\begin{equation}
	\psi = X(x)T(t)~,
\end{equation}
which transforms equation~\ref{eq:kg} into 
\begin{equation}
	X(x)\ddot{T}(t) - c^2T(t)X''(x) + \omega_c^2X(x)T(t) = 0
\end{equation}
and can be separated as
\begin{equation}
	\frac{\omega_c^2T(t) + \ddot{T}(t)}{c^2T(t)} = \frac{X''(x)}{X(x)} = q ~.
\end{equation}
Here, since $X'(x) = ikX(x)$, $q=-k^2$ is the separation constant and we may solve the set of separated equations:
\begin{align}
	X_n''(x) + k_n^2X_n(x) = 0 \nonumber \\
	\ddot{T}_n(t) + (\omega_c^2 + c^2k_n^2)T_n(t) = 0
\end{align}
that can be solved for the complex mode $k=k_r+ik_i$ and written as 
\begin{align}
	X_n = C_{1n} e^{i2n\pi x}e^{ik_i} + C_{2n}e^{-i2n\pi x}e^{-ik_i} \nonumber\\
	T_n = C_{1n}\cos \left (\sqrt{c^2k_n^2+\omega_c^2}t \right ) + C_{2n}\sin \left (\sqrt{c^2k_n^2+\omega_c^2}t \right )~,
\end{align}
where $k_r = 2\pi n; \ \ n=0,\pm1, \pm2, ...$.
If we assume that $\frac{\partial \psi}{\partial t} = 0$, we can write a solution for $\psi$ as the following superposition of modes,
\begin{equation}\label{eq:psi}
	\psi(x,t) = \sum_{n=-N/2}^{N/2} C_ne^{ik_n x}\cos \left (\sqrt{c^2k_n^2+\omega_c^2}t \right ) ~,
\end{equation}
(absorbing all the constants in $C_{1n}$ and $C_{2n}$).
At $t=0$
\begin{equation}
	\psi(x,0) = \sum_{n=-N/2}^{N/2} C_ne^{ik_n x}~,
\end{equation}
and the coefficients can be calculated using a Fourier transform
\begin{equation}
	C_n = \int_{-L}^L\psi_0(x) e^{-ik_{r,n} x}dx = \int_{-L}^L\psi_0(x) e^{-i2\pi n x}dx ~.
\end{equation}

For the initial disturbance, we assume a Gaussian of effective width of twice the Compton wavelength centered at the particle location, $x_p$, as follows. This parameter is chosen here to fit the initial particle PDF solution of our ensemble simulations.
The initial disturbance takes the form
\begin{equation}
	\psi_0(x) = \beta e^{-\left (\frac{x-x_p}{a}\right)^2}~,
\end{equation}
from which the coefficients,  
\begin{align}\label{eq:Cn}
	C_n = \beta \int_{-L}^L  e^{-\left (\frac{x-x_p}{a}\right)^2} e^{-i2\pi n x}dx  
\end{align}
may be calculated.
Finally, by substituting the solution for equation~\ref{eq:Cn} in equation~\ref{eq:psi}, we can express the spatiotemporal evolution of the wave to an initial Gaussian disturbance as 
\begin{align}\label{eq:psi_sol}
	\psi(x,t) = \beta \sum_{n=-N/2}^{N/2} \left \{ e^{i 2\pi n x}\cos \left( \sqrt{c^2k_n^2+\omega_c^2}t \right ) \frac{\sqrt{\pi}a}{2}e^{- \pi^2a^2n^2-i2\pi n x_p} \right . ~ \nonumber \\ \times \left . \left[ erf \left ( \frac{L+x_p}{a}-i\pi n a \right ) + erf \left ( \frac{L-x_p}{a} + i\pi n a \right ) \right ] \right\} ~.
\end{align}

\begin{figure}[t]
\begin{center}       
	\includegraphics[width=1.0\textwidth]{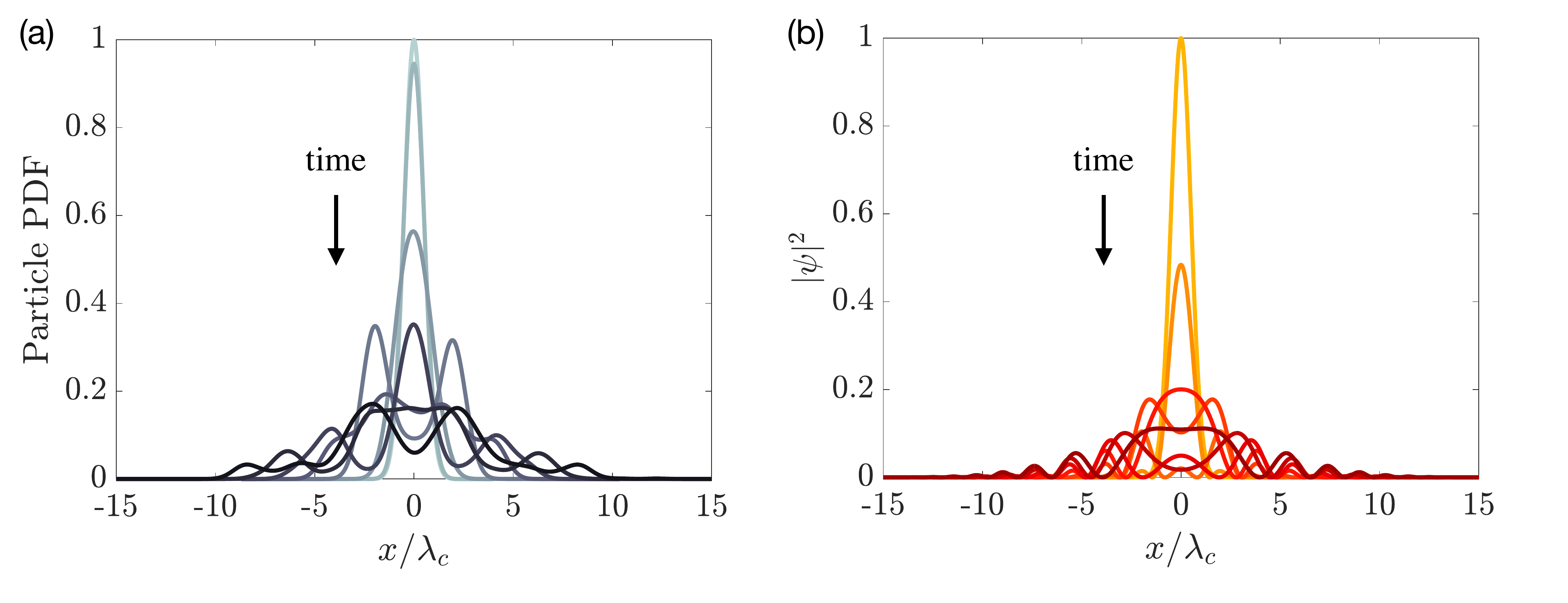}
	\caption{(a) Temporal evolution of particle probability distribution function collected from 1000 simulations (illustrated here in figure~\ref{fig:ensemble}). (b) Normalized analytical solutions for the Gaussian wave response of the time-dependent Klein-Gordon equation. The absolute square value is shown for particle statistics.}
	\label{fig:KG_an_sim}
 \end{center}
\end{figure}

The temporal evolution of the Klein-Gordon wavefield can now be plotted and compared to the evolution of particle PDF in our analogy. 
For the interpretation with Born's rule, the squared absolute value of the wavefield, $\arrowvert \psi \arrowvert^2$, is shown in 
figure~\ref{fig:KG_an_sim}b, and compared to the temporal evolution of the particle PDF extracted from the ensemble simulations (figure~\ref{fig:KG_an_sim}a). 

One may observe a general similarity between the two plots,
although they are not the same.
Initially, the Gaussian distribution is identical between the KG response and the particle PDF.
However, as expected, as time proceeds, the main discrepancies between the two plots emerge at points where the wavefield nullifies. These locations are not allowed in the standard statistical quantum mechanical interpretation. Clearly, this is not the case for the particle PDF in our simulations since all trajectories are generally possible.
Nevertheless, the apparent quantum statistics emerge in our hydrodynamic interpretation through the emergence of coherent structures within the spatially evolving wavefield. 
More specifically, the regions in which particles are less likely to cross (see figure~\ref{fig:ensemble}b) are responsible for the apparent troughs in the PDF of figure~\ref{fig:KG_an_sim}a.

\section{Concluding remarks}
Following our previous work on relativistic pilot-wave theory, we introduced a new deterministic ensemble interpretation based on a hydrodynamically-inspired pilot-wave framework. 
This analysis proposes a plausible mechanism for quantum statistics emerging from the ensemble of multiple uncorrelated deterministic particle trajectories.
Each trajectory is characterized by a quasi-steady speed modulated by inline oscillations at the de Broglie wavelength, which due to nonlinear particle-wave interactions, occasionally changes its direction of motion. 

We show that through the ensemble of particle trajectories, clear, coherent spatiotemporal structures appear at which particles are less likely to cross, similar to previous works on hydrodynamic analogies.
However, the coherent structures in the current framework are comparable to de Broglie wavelength, suggesting a deterministic mechanism for single-particle diffraction on the quantum mechanical scale. 

Finally, we find that the evolution of the particle PDF of this interpretation closely follows the evolution of the square of the absolute-valued Klein-Gordon wave field, suggesting a deterministic approach to Born's rule.  
This is in contrast to the standard Copenhagen interpretation and many-worlds interpretation, in which particles exist nonlocally in all possible locations predicted by quantum statistics.

Although simplified, we hope that this model and its findings may provoke a renewed discussion on realistic pilot waves and the possible resolution of the long sought-after `hidden variables' in the dynamic of quantum particles.

\bibliography{hqftc}
\end{document}